\documentclass{article}

\usepackage{microtype}
\usepackage{graphicx}
\usepackage{subfigure}
\usepackage{amsmath}  
\usepackage{amssymb}  
\usepackage{booktabs} 
\usepackage{multirow} 
\usepackage{enumitem} 
\usepackage{xcolor}   
\usepackage{tikz}     
\usetikzlibrary{arrows.meta,positioning}
\usepackage{caption}  

\definecolor{codebg}{HTML}{F8F8F8}
\definecolor{codeframe}{HTML}{CCCCCC}

\usepackage{listings}
\lstdefinestyle{cstyle}{
  language=C,
  basicstyle=\ttfamily\scriptsize,
  breaklines=true,
  columns=fullflexible,
}

\usepackage{hyperref}

\usepackage[accepted]{mlsys2025}
\makeatletter
\renewcommand{\Notice@String}{}
\makeatother

\newcommand{\tecompile}{\texttt{te-comp}}
\newcommand{\tedialect}{\texttt{te}}
\newcommand{\smlad}{\texttt{SMLAD}}
\newcommand{\rodata}{\texttt{.rodata}}
\newcommand{\bss}{\texttt{.bss}}
\newcommand{\KB}{\,KB}
\newcommand{\MB}{\,MB}

\mlsystitlerunning{te-comp: A Complete MLIR Compiler Generating Self-Contained C Inference Engines for MCUs}

\begin{document}

\twocolumn[
\mlsystitle{te-comp: A Complete MLIR Compiler Generating Self-Contained C Inference Engines for MCUs}

\mlsyssetsymbol{equal}{*}

\begin{mlsysauthorlist}
\mlsysauthor{Van Chan Ngo}{ed}
\end{mlsysauthorlist}

\mlsysaffiliation{ed}{TeComp AI, CA, US}

\mlsyscorrespondingauthor{Van Chan Ngo}{vchngo@gmail.com}

\mlsyskeywords{Machine Learning, MLSys}

\vskip 0.3in

\begin{abstract}

Deploying deep learning on microcontrollers (MCUs) requires fitting both the
model weights and runtime activations within tens to hundreds of kilobytes of
SRAM and Flash.  Existing approaches either use an interpreter with a
pre-written kernel library or combine a Python code generator with hand-written C kernels. 
This incurs dispatch overhead and requires framework dependencies. None produces
self-contained, dependency-free C code specialized to a given model.

In this paper, we present \tecompile{}, the first complete MLIR-based compiler that takes a
standard model in \texttt{.tflite} or \texttt{.safetensors} format and emits
all C source files needed for deployment: the inference engine (kernels,
weight arrays, memory layout) and the RTOS integration layer with zero runtime dependencies.

On 10 MCUNet CNN benchmarks, \tecompile{} outperforms MCUNet TinyEngine by using 38\% less SRAM
(up to 5.9$\times$) and 15\% less Flash, while generating complete C code rather than linking a pre-written kernel library.
To demonstrate, we use \tecompile{} to compile an 8.3M-parameter GPT-Neo transformer
(TinyStories-v2) and deploy it on a \textbf{single-core Cortex-M7 at 160\,MHz}. The full firmware uses 
249\KB{} of 512\KB{} SRAM and 3.62\MB{} of 4\MB{} Flash and achieves approximately \textbf{1\,token/s} interactive text generation via Q4\_0
weight quantization, a stateful int4 KV cache, and vocabulary pruning---a capability no existing MCU inference framework supports.

\end{abstract}

]



\printAffiliationsAndNotice{\mlsysEqualContribution} 

\section{Introduction}\label{sec:intro}

Microcontrollers (MCUs) power billions of embedded devices, from wearable
health monitors and keyword-spotting earbuds to industrial sensors and
autonomous vehicles.  Running deep learning inference directly on these devices
eliminates cloud round-trips, reduces latency to microseconds, and preserves
user privacy.  However, MCUs operate under extreme resource constraints:
typical ARM Cortex-M4/M7 targets offer 128--512\KB{} of SRAM, 512\KB{}--4\MB{}
of Flash, and clock speeds of 160--400\,MHz with no GPU, no virtual memory, and
no dynamic allocation. Existing deployment approaches fall into the following categories, 
each with fundamental limitations.

\textbf{Interpreter runtimes} (TFLite Micro~\cite{david2021tflm},
ExecuTorch~\cite{executorch2024}) load a serialized model and dispatch
operations through a generic interpreter loop.  This is portable but incurs
20--50\KB{} of SRAM overhead from dispatch tables, op resolvers, and runtime
buffer management.  Latency suffers from indirect function calls on every
operator.  Neither framework supports large language model (LLM) inference on
MCUs.

\textbf{Library-based code generators} (MCUNet~\cite{lin2020mcunet,
lin2021mcunetv2}) use a Python script to generate a model-specific dispatch
function that calls into a hand-written C kernel library (33 source files for
MCUNet).  This achieves near-optimal latency through carefully tuned SMLAD-based
kernels, but adding a new operator requires writing new C files, and the
generated code depends on the entire kernel library at link time.

\textbf{General-purpose ML compilers} (Apache TVM~\cite{chen2018tvm},
IREE~\cite{iree}, Glow~\cite{rotem2019glow}) use compiler IRs to optimize
models, but none produces deployable bare-metal code for Cortex-M
microcontrollers.  TVM's Cortex-M support is immature and still requires a
runtime.  IREE and Glow target GPUs and Cortex-A mobile processors.

None of these approaches produces \emph{self-contained, dependency-free C code}
specialized to a specific input model.  We believe that this is the right
abstraction for deployment: the output should be plain C code that firmware 
engineers just drop into any existing project and call with simple APIs without an interpreter, kernel
library, or framework dependencies.

We present \tecompile{}, the first complete MLIR-based compiler that achieves this goal.  
Given a standard model in \texttt{.tflite} for int8 CNNs or \texttt{.safetensors} for transformers, it produces:
\begin{itemize}[nosep]
  \item \texttt{\{model\}\_engine.c/.h}: The complete inference
    engine with all kernels as nested loops, weight arrays in
    \rodata{}, and the activation arena in \bss{}.
  \item \texttt{inference\_task.c} / \texttt{inference\_api.h}: Integration layer (FreeRTOS or bare-metal).
  \item \texttt{CMakeLists.txt}: Drop-in CMake library.
\end{itemize}
Every loop bound, buffer offset, and arithmetic constant is a compile-time
literal specialized to the input model (easy to optimize for machine code).  The generated code has zero external
function calls, not even \texttt{memcpy}. It achieves this through a multi-level MLIR compilation pipeline. A
custom \tedialect{} dialect captures 34 quantized NN operations with all
metadata and composable optimization passes progressively lower and optimize the
IR before emitting C code. Our contributions are:

\begin{enumerate}[nosep]
  \item \textbf{Self-contained C code.}
    Given any input model, \tecompile{} emits all C artifacts needed for deployment: the inference engine, RTOS
    glue code, a public API header, and a CMake library target with zero
    runtime dependencies.  No other tool produces this complete output.
  \item \textbf{LLM inference on Cortex-M7.}
    \tecompile{} compiles an 8.3M-parameter GPT-Neo (TinyStories-v2) to a
    self-contained C engine using 249\KB{} SRAM and 3.62\MB{} Flash
    (measured on hardware), achieving $\sim$1\,token/s interactive
    generation. No existing MCU framework supports this capability.
  \item \textbf{Outperforms MCUNet.} On all 10 CNN benchmarks, \tecompile{} outperforms for both
    peak SRAM (avg 0.62$\times$, up to 5.9$\times$ less) and Flash (avg
    0.86$\times$, up to 32\% less).
  \item \textbf{Numerical accuracy.} \tecompile{} matches TFLite on 10 int8 models with 50-70+ layers by generating identical ground-truth
    label accuracy.
\end{enumerate}

\section{Background}\label{sec:background}

\subsection{MLIR and TOSA}\label{sec:bg-mlir}

MLIR (Multi-Level Intermediate Representation)~\cite{lattner2021mlir} is a
compiler infrastructure that organizes IR into \emph{dialects} --- named
collections of operations, types, and attributes.  A compilation pipeline
progressively lowers operations from high-level dialects to lower-level ones
(e.g., \texttt{linalg} $\to$ \texttt{scf} $\to$ \texttt{arith} $\to$ LLVM),
with each level exposing different optimization opportunities.

TOSA (Tensor Operator Set Architecture)~\cite{tosa2021} is an MLIR dialect
that provides a portable, quantized tensor operation set.  It serves as the
entry point for \tecompile{}: a Python converter translates \texttt{.tflite}
int8 models into TOSA MLIR, preserving quantization parameters (zero points,
scales) as tensor operands.  TOSA represents operations like \texttt{tosa.conv2d}
with explicit quantization semantics, making it suitable as a stable input
format for quantized model compilation.

\subsection{Cortex-M DSP Instructions}\label{sec:bg-dsp}

ARM Cortex-M4 and M7 processors include DSP extensions with packed-arithmetic
instructions.  The \smlad{} (Signed Multiply Accumulate Dual) instruction
performs two 16-bit multiplications and accumulates both results into a 32-bit
accumulator in a single cycle:
\begin{equation}
  \texttt{SMLAD}(a, b, c) = c + (a_\text{lo} \times b_\text{lo}) + (a_\text{hi} \times b_\text{hi})
\end{equation}
where $a_\text{lo/hi}$ and $b_\text{lo/hi}$ are the lower and upper 16-bit
halves of 32-bit registers.  Combined with \texttt{SXTB16} (sign-extend two
bytes to halfwords) and \texttt{PKHBT} (pack halfwords) instructions, this enables
efficient int8 MAC (Multiply-Accumulate) operations, two MACs per cycle
instead of one. Standard C compilers (GCC, LLVM) do \emph{not} emit \smlad{} from idiomatic
\texttt{for-loop} code.  Auto-vectorization targets NEON (Cortex-A) or MVE
(Cortex-M55+), not the DSP extension.  Generating \smlad{}-based kernels
requires either inline assembly, compiler intrinsics, or as in
\tecompile{}, a compiler IR that directly models these patterns.

\subsection{MCUNet TinyEngine}\label{sec:bg-mcunet}

MCUNet~\cite{lin2020mcunet,lin2021mcunetv2} is a co-designed neural
architecture search and inference engine for microcontrollers.  The
TinyEngine inference library uses a \emph{hybrid} code generation approach:
\begin{itemize}[nosep]
  \item A Python code generator emits a model-specific dispatch function that calls kernel routines in a fixed order.
  \item Kernel implementations are a hand-written C library (33 source files)
    selected by convolution type and shape (pointwise 1$\times$1, depthwise,
    general).
  \item Memory management uses a static arena with a memory scheduler that
    plans buffer offsets at compile time.
\end{itemize}

TinyEngine achieves state-of-the-art inference latency on Cortex-M through
partial im2col with \smlad{}-based kernels, in-place depthwise
convolution (MCUNetV2), and operator specialization.  However, the kernel
library is not generated (it must be linked in its entirety), and adding new
operators requires writing new C files.  The code generator uses Python string
templates rather than a compiler IR, making optimizations difficult to compose
and verify.

\tecompile{} adopts MCUNet's key algorithmic innovations (in-place depthwise,
partial im2col, operator specialization) and re-implements them as composable
MLIR passes that generate all kernel code from IR. Thus, it eliminates the 
need for a pre-written library.

\section{Compiler Design}\label{sec:design}

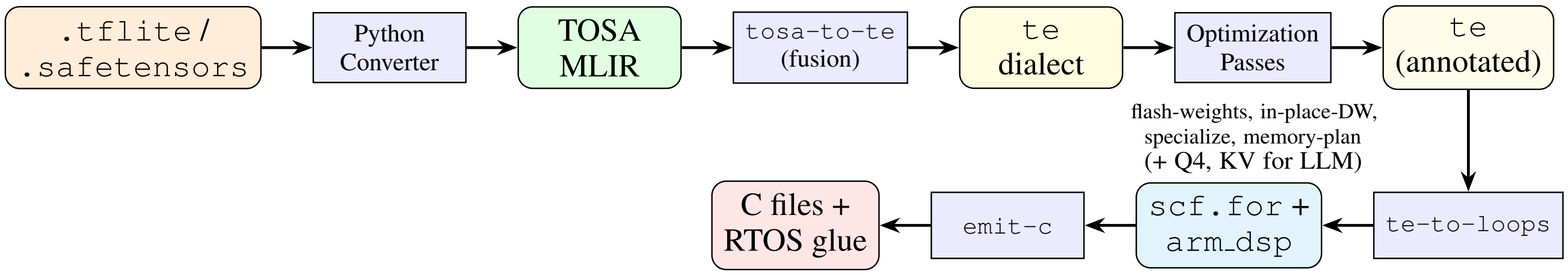
\begin{figure*}[t]
  \centering
  \resizebox{\textwidth}{!}{%
  \begin{tikzpicture}[
    node distance=0.4cm and 0.5cm,
    block/.style={rectangle, draw, rounded corners, minimum height=0.8cm,
                  minimum width=1.6cm, align=center, font=\footnotesize},
    pass/.style={rectangle, draw, fill=blue!8, minimum height=0.65cm,
                 minimum width=1.5cm, align=center, font=\scriptsize},
    arr/.style={-{Stealth[length=2mm]}, thick},
  ]
    \node[block, fill=orange!15] (input) {\texttt{.tflite} /\\\texttt{.safetensors}};
    \node[pass, right=of input] (conv) {Python\\Converter};
    \node[block, fill=green!12, right=of conv] (tosa) {TOSA\\MLIR};
    \node[pass, right=of tosa] (tosa2te) {\texttt{tosa-to-te}\\(fusion)};
    \node[block, fill=yellow!15, right=of tosa2te] (te) {\tedialect{}\\dialect};
    \node[pass, right=of te] (opt) {Optimization\\Passes};
    \node[block, fill=yellow!10, right=of opt] (teopt) {\tedialect{}\\(annotated)};

    \node[font=\tiny, below=0.08cm of opt, text width=2.5cm, align=center]
      (optlabel) {
      flash-weights, in-place-DW,\\specialize, memory-plan\\{\scriptsize(+ Q4, KV for LLM)}
    };

    \node[pass, below=1.0cm of teopt] (lower) {\texttt{te-to-loops}};
    \node[block, fill=cyan!10, left=of lower] (loops) {\texttt{scf.for} +\\\texttt{arm\_dsp}};
    \node[pass, left=of loops] (emit) {\texttt{emit-c}};
    \node[block, fill=red!10, left=of emit] (cfiles) {C files +\\RTOS glue};

    \draw[arr] (input) -- (conv);
    \draw[arr] (conv) -- (tosa);
    \draw[arr] (tosa) -- (tosa2te);
    \draw[arr] (tosa2te) -- (te);
    \draw[arr] (te) -- (opt);
    \draw[arr] (opt) -- (teopt);

    \draw[arr] (teopt) -- (lower);

    \draw[arr] (lower) -- (loops);
    \draw[arr] (loops) -- (emit);
    \draw[arr] (emit) -- (cfiles);
  \end{tikzpicture}%
  }
  \caption{A model file is converted to
    TOSA MLIR, lowered to the \tedialect{} dialect with op fusion, optimized
    by composable passes, lowered to loops with DSP intrinsics, and emitted as
    self-contained C source files.}
  \label{fig:pipeline}
\end{figure*}

Figure~\ref{fig:pipeline} shows the \tecompile{} compilation pipeline.  The
compiler processes both CNN (\texttt{.tflite}) and transformer
(\texttt{.safetensors}) models through a shared multi-level IR, with different
optimization pass configurations selected automatically based on model type.

\subsection{The \tedialect{} Dialect}\label{sec:te-dialect}

The \tedialect{} dialect defines 34 operations for quantized neural network
inference, organized into four categories (Table~\ref{tab:te-ops}). 
The Appendix provides the complete list of operations and their descriptions. Each operation carries all metadata required for code generation as MLIR
attributes:
\begin{itemize}[nosep]
  \item \emph{Quantization}: input/weight/output zero points, per-channel
    multiplier and shift arrays, output clamp bounds.
  \item \emph{Layout}: strides, padding, dilation, kernel dimensions.
  \item \emph{Memory}: arena offset that is populated by the memory planner.
  \item \emph{Specialization hints}: kernel variant (pointwise, depthwise, general),
    unrolling factors, in-place flag.
\end{itemize}

\begin{table}[t]
  \centering
  \caption{The 34 operations in the \tedialect{} dialect.}
  \label{tab:te-ops}
  \small
  \begin{tabular}{@{}l l@{}}
    \toprule
    \textbf{Category} & \textbf{Operations} \\
    \midrule
    CNN      & \texttt{conv2d}, \texttt{depthwise\_conv2d}, \\
    (22 ops) & \texttt{transpose\_conv2d}, \texttt{fully\_connected}, \\
             & \texttt{avg\_pool2d}, \texttt{max\_pool2d}, \\
             & \texttt{add}, \texttt{mul}, \texttt{max}, \texttt{min}, \\
             & \texttt{clamp}, \texttt{rescale}, \texttt{softmax}, \\
             & \texttt{sigmoid}, \texttt{tanh}, \texttt{hard\_swish}, \\
             & \texttt{leaky\_relu}, \texttt{reshape}, \texttt{transpose}, \\
             & \texttt{concat}, \texttt{pad}, \texttt{upsample} \\
    \midrule
    RNN (2)  & \texttt{svdf}, \texttt{lstm} \\
    \midrule
    LLM      & \texttt{matmul\_q4}, \texttt{gather}, \texttt{layer\_norm}, \\
    (9 ops)  & \texttt{gelu}, \texttt{rope}, \texttt{kv\_cache\_write}, \\
             & \texttt{kv\_cache\_attention}, \texttt{kv\_slide}, \\
             & \texttt{pos\_embed} \\
    \midrule
    Special  & \texttt{se\_block} (squeeze-and-excitation) \\
    \bottomrule
  \end{tabular}
  \begin{flushleft}
    \scriptsize All operations are prefixed with \texttt{te.}
    (e.g., \texttt{te.conv2d}).
  \end{flushleft}
\end{table}

A key design decision is introducing a \emph{custom} dialect rather than
lowering TOSA directly to loops.  TOSA operations are too coarse for
optimization --- a \texttt{tosa.conv2d} followed by \texttt{tosa.rescale}
and \texttt{tosa.clamp} must be fused into a single \texttt{te.conv2d} before
memory planning can reason about activation lifetimes.  The \tedialect{}
dialect provides a level where optimization passes can attach annotations
(arena offsets, specialization hints, in-place flags) that are consumed by the
kernel lowering pass.

\subsection{Operation Fusion}\label{sec:fusion}

\tecompile{} fuses multi-op TOSA patterns into single \tedialect{}
operations, reducing both memory traffic and intermediate buffer
allocations.  Three classes of fusion are applied:
\begin{itemize}[nosep]
  \item \emph{Quantized convolution}: A \texttt{tosa.conv2d} followed by \texttt{tosa.rescale} (per-channel
  requantization) and \texttt{tosa.clamp} (activation bounds) fuses into a
  single \texttt{te.conv2d} with multiplier/shift arrays and
  clamp bounds as attributes.  The same pattern applies to
  \texttt{depthwise\_conv2d} and \texttt{fully\_connected}.
  \item \emph{Five-op normalization}: The converter decomposes $\texttt{LayerNorm}(x, \gamma, \beta)$ into 11~TOSA operations since TOSA has no native LayerNorm. 
  \tecompile{} fuses them into a single \texttt{te.layer\_norm}. With $2L{+}1$ normalizations in an $L$-layer transformer
  ($L{=}8$ for TinyStories), this eliminates $17 \times 10 = 170$ intermediate
  tensors from the memory plan.
  \item \emph{Six-op activation}: The \texttt{GELU} activation decomposes into a 9-TOSA-operation sequence. \tecompile{} fuses this entire sequence into a single
  \texttt{te.gelu}, eliminating eight intermediate buffers. With one \texttt{GELU} per transformer layer, the $L{=}8$ model saves $8 \times 8 = 64$ intermediate tensors.
\end{itemize}
All fusion patterns are expressed as MLIR rewrite rules and verified by
the MLIR type system, unlike the Python string templates (MCUNet) or
hand-written kernel selection tables (CMSIS-NN) used by prior work.

\subsection{Static Arena Memory Planner}\label{sec:memory-planner}

The memory planner maps every activation tensor including model I/O to offsets within a byte array at compile
time.  Operating on the \tedialect{} IR where all shapes are statically
known, it produces a deterministic, zero-overhead layout with no runtime
allocator.  The algorithm proceeds in five phases:
\begin{itemize}[nosep]
  \item \emph{Lifetime analysis}: Assign each tensor a
    $[\textit{def},\;\textit{last\_use}]$ interval and byte size.
  \item \emph{In-place aliasing}: Tensors from in-place depthwise ops or
    zero-copy reshapes inherit their input's offset; stride-2 ops further
    \emph{shrink} the root buffer after execution, freeing the tail.
  \item \emph{Patch-size reduction}: Substitute smaller patched sizes when
    all consumers are patch-inference operations (Section~\ref{sec:specialize}).
  \item \emph{FirstFit-decreasing packing}: Sort tensors by size, place
    each at the lowest non-conflicting offset.
  \item \emph{Annotation}: Attach \texttt{arena\_offset} attributes to
    every operation result and record \texttt{te.peak\_mem} attribute on the function.
\end{itemize}
Following the MCUNet design~\cite{lin2020mcunet}, model I/O lives inside
the arena; the engine exposes zero-copy accessor APIs so the caller reads and
writes directly at the planned offsets.  In addition to the arena, the
planner computes three scratch buffers: \emph{im2col} (two \texttt{i16}
columns for the SMLAD kernel), \emph{depthwise} (one padded spatial plane),
and \emph{Q8 activation} (for \texttt{te.matmul\_q4}) whose sizes are recorded
as function attributes.  Total peak SRAM equals arena~+~scratch.

\subsection{CNN and LLM Pipelines}\label{sec:cnn-llm}

Both model types enter the compiler as TOSA MLIR but follow different
optimization pass configurations:
\begin{itemize}[nosep]
  \item \textbf{CNN}: tosa-to-te $\to$ flash-weights $\to$
    in-place-DW $\to$ specialize $\to$ memory-plan $\to$
    te-to-loops $\to$ arm-dsp-to-llvm $\to$ emit-c.
  \item \textbf{LLM}: flash-weights $\to$ kv-cache $\to$
    quantize-weights $\to$ parallel-matmul $\to$ tosa-to-te $\to$
    memory-plan $\to$ te-to-loops $\to$ emit-c.
\end{itemize}
The model type is auto-detected from the input file format: \texttt{.tflite}
triggers the CNN pipeline, \texttt{.safetensors} with a transformer config
triggers the LLM pipeline.  The same \tecompile{} binary handles both.

\section{Optimization Passes}\label{sec:optimization}

We now describe the CNN-specific and LLM-specific
passes.  Each is a standalone MLIR transformation that can be composed,
reordered, or disabled independently. The Appendix provides the complete list of passes implemented by \tecompile{} and their
descriptions.

\subsection{Flash-Resident Weights}\label{sec:flash-weights}

The \texttt{--te-flash-weights} pass marks all weight constant tensors as
flash-resident by annotating them with a \texttt{flash\_resident} attribute.
During C emission, these tensors are emitted as \texttt{static const}
arrays in \rodata{}, which the linker places in Flash memory.  This removes
weights from SRAM entirely. Thus, for MCUNet models with 0.37--1.73M parameters,
this saves 370\KB{}--1.73\MB{} of SRAM that would otherwise be consumed at
startup.

\subsection{In-Place Depthwise Convolution}\label{sec:inplace-dw}

Depthwise separable convolutions process each input channel independently.
When stride${}=1$ with same-padding, the output has the same spatial
dimensions as the input.  The \texttt{--te-in\allowbreak{}-place\allowbreak{}-depthwise}
pass sets an \texttt{in\_place} flag on qualifying depthwise operations,
instructing the memory planner to alias the output tensor to the same arena
offset as the input.

This optimization, first proposed in MCUNetV2~\cite{lin2021mcunetv2}, halves
the activation memory for every depthwise layer.  In \tecompile{}, it is
implemented as a composable MLIR pass that modifies tensor lifetime annotations
--- the memory planner extends the input's lifetime to cover the output's last
use, and the kernel lowering pass generates an in-place write loop.

\subsection{Operator Specialization}\label{sec:specialize}

The \texttt{--te-operator-specialize} pass annotates convolution operations
with kernel variant hints consumed by the lowering pass:
\begin{itemize}[nosep]
  \item \textbf{Pointwise 1$\times$1}: Skips the $k_h, k_w$ loops entirely
    (2 fewer nested loops per operation), converting the convolution to a
    matrix multiply over spatial positions.
  \item \textbf{Depthwise}: Annotated for a channel-first loop structure
    that enables in-place operation and avoids im2col.
  \item \textbf{Channel unrolling}: Operations with small channel counts
    receive unrolling hints to eliminate loop overhead.
\end{itemize}

\subsection{Memory Planning Interaction}\label{sec:memory-plan}

The \texttt{--te-memory-plan} pass invokes the static arena memory planner (Section~\ref{sec:memory-planner}). It is the
downstream consumer of the CNN optimization passes above.  In-place
depthwise aliasing, I/O-in-arena packing, and FirstFit-decreasing bin
packing combine to reduce peak SRAM by 27--83\% on MCUNet models compared
to naive sequential allocation (Table~\ref{tab:combined}).  During C
emission, the arena is realized as a single
\texttt{static int8\_t buffer[N]} array in \bss{}, where $N$ is the peak
arena size reported by the planner.

\subsection{LLM-Specific Passes}\label{sec:llm-passes}

For transformer models such as TinyStories-v2, \tecompile{} provides the following additional passes.

\textbf{Q4\_0 weight quantization}: The \texttt{--te-quantize\allowbreak{}-weights} pass quantizes fp32 weight matrices to 4-bit using a block-wise format
(32-element blocks, 4.5 bits/weight effective).  Thus, it achieves $\sim$7$\times$ Flash
compression.  The pass reports per-layer RMSE, SNR, and maximum error at
compile time, warning if any layer drops below a configurable SNR threshold.

\textbf{Int4 KV cache}: The \texttt{--te-kv-cache} pass inserts quantization and
dequantization operations around KV cache writes and reads, reducing SRAM
usage by 8$\times$ compared to fp32 storage.  The cache format (int4/int8/fp32)
is configurable.

\textbf{Vocabulary pruning}: The \texttt{--prune-vocab} pass prunes the embedding
table and LM head to a corpus-observed token set.  For TinyStories-v2 with
14.6K kept tokens, the embedding shrinks from 50,257$\times$128 to
14,632$\times$128, saving $\sim$71\% of embedding Flash.

\textbf{Parallel matmul}: The \texttt{--te-parallel-matmul} pass splits matrix
multiplications across multiple cores via row partitioning, generating
per-core loop bounds.

\subsection{Automated Optimization Suggestion}\label{sec:auto-opt}

An AOT (Ahead-of-Time) analysis pass is included in \tecompile{}, \texttt{--te-suggest-optimizations}. 
It inspects the TOSA IR and
produces a prioritized recommendation report.  It extracts the input model
signature (op types, weight formats, attention dimensions), estimates
memory requirements, and cross-references against user-supplied target
constraints (SRAM, Flash, core count) to produce a prioritized list of
applicable passes (\texttt{HIGH}, \texttt{MEDIUM}, \texttt{LOW}, \texttt{SKIP}) with estimated savings and
accuracy impact.  An \texttt{--auto-optimize} flag applies all
non-\texttt{SKIP} recommendations automatically.

\section{Code Generation}\label{sec:codegen}

After optimization, \tecompile{} lowers the annotated \tedialect{} IR to
loops and emits complete C source files with the \texttt{--emit-c} backend.  This section describes the kernel
lowering strategy and the output structure.

\subsection{Kernel Lowering}\label{sec:kernel-lowering}

Each \tedialect{} operation is lowered to nested \texttt{scf.for} loops
operating on \texttt{memref} buffers.  For \texttt{te.conv2d}, the lowering
produces a 5-deep loop nest (OH, OW, OC, KH, KW). The IC dimension is folded into the \smlad{} paired MAC and the actual output includes bounds checking and per-channel requantize.

\begin{figure}[t]
\begin{lstlisting}[style=cstyle, numbers=left, numbersep=4pt, backgroundcolor=\color{codebg}]
// Conv2D: N=1, OH=48, OW=48, OC=32, KH=3, KW=3, IC=3
for (int oh = 0; oh < 48; oh++) {
  for (int ow = 0; ow < 48; ow++) {
    for (int oc = 0; oc < 32; oc += 2) {
      int32_t acc0 = 0, acc1 = 0;
      for (int kh = 0; kh < 3; kh++) {
        for (int kw = 0; kw < 3; kw++) {
          // Partial im2col: pack 2 i8 inputs
          // into i16 pairs, then SMLAD
          int32_t a = SXTB16(input[...]);
          int32_t b = SXTB16(weight[...]);
          acc0 = SMLAD(a, b, acc0);
          // ... (2nd output channel)
        }
      }
      // Per-channel requantize + clamp
      out[oh][ow][oc] = clamp(
        (acc0 * mult[oc] + round) >> shift[oc]
        + out_zp, -128, 127);
    }
  }
}
\end{lstlisting}
\caption{Simplified generated Conv2D kernel showing the lowered nested \texttt{scf.for} loops.}
\label{lst:conv2d}
\end{figure}

For pointwise 1$\times$1 convolutions, the $k_h$ and
$k_w$ loops are eliminated entirely (Section~\ref{sec:specialize}). 
Every loop bound is a compile-time integer literal because \tecompile{} knows the exact model shape. 
This makes the generated code easy to optimize for standard C compilers.

The two-column \smlad{} GEMM processes 2 output pixels $\times$ 2 output
channels per inner iteration, achieving 2 MAC operations per \smlad{}
instruction on ARM Cortex-M.

\subsection{C Code Emission}\label{sec:c-emit}

The EmitC backend produces the following files:
\begin{itemize}[nosep]
  \item \emph{Inference engine source}: The complete engine with all kernels as nested loops, weight arrays in \rodata{} section and activation arena in \bss{} section. The arena is hoisted to file scope so that the accessor functions can return pointers into it.
  \item \emph{Integration layer source}: The complete set of APIs to initialize, run, de-initialize the inference engine, set model input and get model output with zero copy pattern.
  \item \emph{Manifest}: A \texttt{.json} file with machine-readable metadata such as arena size, weights, layer details, and MACs.
\end{itemize}

\subsection{Integration Layer}\label{sec:rtos}

The compiler generates integration layer code from templates in two configurations: RTOS task or bare-metal.
\begin{itemize}[nosep]
  \item \texttt{inference\_task.c}: This implements a FreeRTOS task or bare-metal wrapper. It contains zero static I/O buffers because the wrapper delegates directly to the engine's arena accessor functions.
  \item \texttt{inference\_api.h}: This declares APIs to interface with the inference engine with zero-copy pattern.
  \item \texttt{CMakeLists.txt}: This defines a drop-in CMake library target.
\end{itemize}

Integration into an existing firmware project requires three lines if the project uses a CMake build system.
\begin{lstlisting}[style=cstyle, numbers=left, numbersep=4pt, backgroundcolor=\color{codebg}]
// CMakeLists.txt
add_subdirectory(te_generated)
target_link_libraries(firmware PRIVATE te_generated)

// Application code (zero-copy)
inference_init(NULL);
memcpy(inference_get_input(), sensor_data,
       INPUT_SIZE_BYTES);        // write into arena
inference_run();                 // in-place, no args
const int8_t *out = inference_get_output();
                                 // read from arena
\end{lstlisting}

\noindent
This matches MCUNet's I/O APIs with zero-copy pattern. The caller's sensor data goes directly into the arena; the result
is read in-place without any \texttt{memcpy} inside the engine.  The entire I/O and activation storage lives in the engine's
single arena buffer.

The generated code compiles with any ARM cross-compiler (\texttt{arm-none-eabi-gcc},
\texttt{armclang}, IAR).  No LLVM, no Python, no framework headers are needed
at the firmware build stage.

\section{Evaluation}\label{sec:eval}

All benchmark scripts, model files, test inputs, and instructions to
reproduce every result in this section are publicly available
at~\cite{tecompbenchmarks}.

\subsection{Experimental Setup}\label{sec:setup}

\textbf{Models.}
We evaluate on all 10 MCUNet benchmark CNN models (7 ImageNet, 3 Visual Wake
Words).  The MCUNet TFLite files are byte-identical to the official releases from the MCUNet
repository~\cite{lin2020mcunet}. For LLM evaluation, we compile TinyStories-v2, GPT-Neo, 8.3M parameters~\cite{eldan2023tinystories}.

\textbf{Metrics.}
SRAM is the peak memory reported by the memory planner: arena (which
includes model input, activations, and output via \textit{FirstFit} packing)
plus scratch buffers (im2col, depthwise pad).  With zero-copy I/O,
the engine allocates no staging buffers (the caller writes input
directly into the arena and reads output in-place) so the arena
is the total SRAM footprint.  This is directly comparable to
MCUNet which also packs I/O inside the allocator and
exposes zero-copy APIs.
Flash is the \texttt{.text} section (code + weights in \rodata{}).
Both verified with \texttt{arm-none-eabi-size} on the final ELF binary.
For the LLM deployment, memory is measured on the complete deployed
firmware (engine + tokenizer + vocabulary remap + driver + startup code).

\textbf{Baselines.}
MCUNet TinyEngine numbers are from the official MCUNet GitHub
repository~\cite{lin2020mcunet}. SRAM and Flash are profiled with their tooling.

\textbf{Compile time.}
\tecompile{} compiles CNN models in 1--5 seconds and LLM models in 5--15
seconds on a modern x86 workstation (single-threaded, including Python
converter, MLIR passes, and C emission).  This is fast enough for interactive
development with an IDE plugin.

\subsection{CNN: te-comp vs.\ MCUNet TinyEngine}\label{sec:eval-mcunet}

Table~\ref{tab:combined} presents the full comparison on all 10 MCUNet
models for both peak SRAM and Flash.

\begin{table*}[t]
  \centering
  \caption{Peak SRAM and Flash of te-comp vs.\ MCUNet TinyEngine on 10 benchmark models. MCUNet numbers from~\cite{lin2020mcunet}.}
  \label{tab:combined}
  \resizebox{\textwidth}{!}{%
  \begin{tabular}{@{}l r r r r r r r r@{}}
    \toprule
    \textbf{Model} & \textbf{MACs} & \textbf{\#Params} &
    \textbf{MCU SRAM (KB)} & \textbf{te SRAM (KB)} & \textbf{SRAM ratio} &
    \textbf{MCU Flash (KB)} & \textbf{te Flash (KB)} & \textbf{Flash ratio} \\
    \midrule
    mcunet-in0      & 6.4M   & 0.75M & 266 & \textbf{45}  & \textbf{0.17$\times$} & 889  & \textbf{823}  & 0.93$\times$ \\
    mcunet-in1      & 12.8M  & 0.64M & 307 & \textbf{90}  & \textbf{0.29$\times$} & 992  & \textbf{704}  & 0.71$\times$ \\
    mcunet-in2      & 67.3M  & 0.73M & 242 & \textbf{200} & 0.83$\times$ & 878  & \textbf{816}  & 0.93$\times$ \\
    mcunet-in3      & 81.8M  & 0.74M & 293 & \textbf{242} & 0.83$\times$ & 897  & \textbf{816}  & 0.91$\times$ \\
    mcunet-in4      & 125.9M & 1.73M & 456 & \textbf{400} & 0.88$\times$ & 1876 & \textbf{1823} & 0.97$\times$ \\
    mbv2-w0.35      & 23.5M  & 0.75M & 308 & \textbf{283} & 0.92$\times$ & 862  & \textbf{820}  & 0.95$\times$ \\
    proxyless-w0.3  & 38.3M  & 0.75M & 292 & \textbf{242} & 0.83$\times$ & 892  & \textbf{824}  & 0.92$\times$ \\
    mcunet-vww0     & 6.0M   & 0.37M & 146 & \textbf{56}  & \textbf{0.38$\times$} & 617  & \textbf{423}  & \textbf{0.69$\times$} \\
    mcunet-vww1     & 11.6M  & 0.43M & 162 & \textbf{87}  & \textbf{0.54$\times$} & 689  & \textbf{496}  & 0.72$\times$ \\
    mcunet-vww2     & 55.8M  & 0.64M & 311 & \textbf{162} & \textbf{0.52$\times$} & 897  & \textbf{734}  & 0.82$\times$ \\
    \midrule
    \textbf{Average} & & & & & \textbf{0.62$\times$} & & & \textbf{0.85$\times$} \\
    \bottomrule
  \end{tabular}%
  }
\end{table*}

\textbf{SRAM.}
\tecompile{} uses less SRAM than MCUNet on all 10 models, with an average
ratio of 0.62$\times$ (38\% less).  The largest improvement is on
\texttt{mcunet-in0}: 266\KB{} $\to$ 45\KB{} (5.9$\times$ reduction).
The gains are largest on small-input models (\texttt{mcunet-in0},
\texttt{vww0}, \texttt{vww1}, and \texttt{vww2}) where the memory planner can more aggressively overlap
activation buffers, and in-place depthwise convolution eliminates a larger
fraction of peak memory.

\textbf{Flash.}
\tecompile{} uses less Flash on all 10 models, with an average ratio of
0.85$\times$ (15\% less).  The savings come from generating only the kernel
code needed for the specific model, whereas MCUNet links its entire 33-file
kernel library.  The largest saving is on \texttt{mcunet-vww0}: 617\KB{}
$\to$ 423\KB{} (31\% less).

\subsection{Numerical Accuracy}\label{sec:accuracy}

We verify accuracy by comparing \tecompile{}'s generated C engine against
the TFLite reference interpreter (v2.14) on all 10 int8 models, using 10
real test samples per model from Imagenette~\cite{imagenette},
MCUNet's VWW minival~\cite{lin2020mcunet}, and standard test sets.
The generated engine is compiled with \texttt{gcc~-O2} on the host with \texttt{target=generic} and compared element-wise against TFLite.
During development, we identified and fixed five numerical bugs in the
TFLite-to-TOSA conversion (requantize rounding, Pad$+$Conv spatial mapping,
negative per-channel scales, softmax approximation, average-pool rounding).
Table~\ref{tab:accuracy} presents the results after these corrections.

\begin{table*}[t]
  \centering
  \caption{Numerical accuracy of \tecompile{} vs.\ TFLite reference
    interpreter on 10 int8 models (10 real test samples per model).
    Max LSB = largest element-wise absolute difference.
    Elem.\ err = fraction of output elements differing by $\geq 1$\,LSB.
    TFL$\rightarrow$GT and Eng$\rightarrow$GT = number of inputs
    (out of 10) where TFLite / \tecompile{} predict the correct
    ground-truth label.}
  \label{tab:accuracy}
  \small
  \begin{tabular}{@{}l r r r r r r r@{}}
    \toprule
    \textbf{Model} & \textbf{Out} & \textbf{Tens.} &
    \textbf{Max LSB} & \textbf{RMSE} &
    \textbf{Elem.\ err} &
    \textbf{TFL$\rightarrow$GT} & \textbf{Eng$\rightarrow$GT} \\
    \midrule
    \multicolumn{8}{@{}l}{\textit{Deep mobile architectures (50--70+ layers)}} \\
    mcunet-320kb-vww     & 2     & 353 & 2          & 1.118 & 65.0\% & 9/10 & 9/10 \\
    mcunet-10fps-vww     & 2     & 225 & 3          & 1.449 & 75.0\% & 9/10 & 9/10 \\
    mcunet-5fps-vww      & 2     & 241 & 3          & 1.449 & 85.0\% & 9/10 & 9/10 \\
    mbv2-w0.35-r144      & 1000  & 278 & 7          & 1.403 & 70.5\% & \textbf{10/10} & \textbf{10/10} \\
    mcunet-256kb-in      & 1000  & 289 & 7          & 1.608 & 74.0\% & \textbf{10/10} & \textbf{10/10} \\
    mcunet-5fps-in       & 1000  & 225 & 7          & 1.625 & 75.0\% & \textbf{10/10} & \textbf{10/10} \\
    proxyless-w0.3-r176  & 1000  & 326 & 7          & 1.519 & 73.6\% & \textbf{10/10} & \textbf{10/10} \\
    mcunet-10fps-in      & 1000  & 241 & 9          & 1.928 & 78.2\% & 9/10  & 9/10 \\
    mcunet-320kb-in      & 1000  & 289 & 9          & 1.412 & 69.6\% & \textbf{10/10} & \textbf{10/10} \\
    mcunet-512kb-in      & 1000  & 273 & 9          & 1.643 & 75.2\% & \textbf{10/10} & \textbf{10/10} \\
    \bottomrule
  \end{tabular}
\end{table*}

The ten deep mobile architectures (50--70+ layers) exhibit 2--9\,LSB
maximum error from accumulated rounding through $O(50)$ requantize
operations. Each operation is within tolerance of 1--2\,LSB but
compounding across the network.  Despite element-level differences,
ground-truth label accuracy is identical to TFLite on all 10
MCUNet models (9/10 or 10/10 on both engines for every model).

\subsection{LLM: TinyStories-v2 on Cortex-M7}\label{sec:eval-llm}

\textbf{Memory.}
\tecompile{} compiles TinyStories-v2 (GPT-Neo, 8.3M parameters, 8 layers,
$d_\text{model}=128$, 16 attention heads) to a self-contained C inference
engine for a single-core ARM Cortex-M7 at 160\,MHz (NXP S32K344:
512\KB{} SRAM, 4\MB{} Flash, \$25 automotive MCU).

Table~\ref{tab:llm-mem} shows the memory breakdown measured on the final ELF binary
deployed to the S32K3X4EVB-T172 evaluation board. The complete firmware including model weights, tokenizer, vocabulary remap table, 
inference engine, NXP RTD drivers, and startup code uses 3.62\MB{} of Flash (90.5\% of 4\MB{}) and
249\KB{} of SRAM (48.5\% of 512\KB{}), leaving headroom for application logic.

\begin{table}[t]
  \centering
  \caption{Memory usage of TinyStories-v2 (pruned 14.6K vocabulary) on S32K344 MCU.}
  \label{tab:llm-mem}
  \small
  \begin{tabular}{@{}l r l@{}}
    \toprule
    \textbf{Component} & \textbf{Size} & \textbf{Section} \\
    \midrule
    Q4\_0 weights + pruned embedding        & 3.02\MB{} & \texttt{.rodata} \\
    BPE tokenizer blob                      & 230\KB{} & \texttt{.rodata} \\
    Generated engine + firmware code        & 382\KB{} & \texttt{.text} \\
    \midrule
    \textbf{Total Flash (\texttt{.text})}   & \textbf{3.62\MB{}} & \textbf{Fits 4\MB{}} \\
    \midrule
    Int4 KV cache + scales                  & 108\KB{} & \texttt{.bss} \\
    Activation arena                        & 58\KB{} & \texttt{.bss} \\
    Tokenizer + sampler + UART buffers      & 83\KB{} & \texttt{.bss} \\
    \midrule
    \textbf{Total SRAM (\texttt{.bss})}     & \textbf{249\KB{}} & \textbf{Fits 512\KB{}} \\
    \bottomrule
  \end{tabular}
\end{table}

\textbf{Throughput.}
On the S32K3X4EVB-T172 board (single Cortex-M7 core at 160\,MHz), the
generated engine achieves approximately \textbf{1\,token/s} for
autoregressive text generation (a single inference run generating one token takes approximately \texttt{1150ms}). 
Each token requires $\sim$16.6M Q4
multiply-accumulate operations through 8 transformer layers.  The engine
runs bare-metal with no OS. A video demonstration of interactive
multi-turn conversation on the physical board is available at~\cite{tecompdemo}.

\textbf{KV Cache.}
The compiler's \texttt{--te-kv-cache} pass detects the attention pattern
in the TOSA IR (Q/K/V projections $\to$ reshape to heads $\to$
matmul $\to$ softmax $\to$ matmul) and replaces it with stateful
\texttt{te.kv\_cache\_write} and \texttt{te.kv\_cache\_attention} operations.
These lower to C code that maintains a persistent int4-quantized KV cache
with a runtime position counter, enabling autoregressive generation where
each call processes one token and attends over all previous tokens.
The position counter auto-increments in engine run function and resets
in engine initialization function, providing correct multi-turn conversation.

\section{Related Work}\label{sec:related}

\textbf{MCUNet TinyEngine}~\cite{lin2020mcunet,lin2021mcunetv2}.
Closest to our work. It uses neural architecture search (NAS) to find
MCU-friendly models and a Python code generator + hand-written C kernel
library for deployment.  \tecompile{} replaces both the Python generator and
the kernel library with a proper MLIR compiler.  We adopt MCUNet's key
optimizations (in-place depthwise, partial im2col, operator specialization)
and implement them as composable, verifiable MLIR passes.  Our evaluation
shows \tecompile{} outperforms MCUNet on all 10 benchmark models for
both SRAM and Flash usage.

\textbf{TFLite Micro}~\cite{david2021tflm}.
The standard interpreter-based MCU runtime. It supports a wide range of
operators via an op resolver that dispatches to pre-registered kernels.
The interpreter overhead (dispatch tables, tensor arena management, op
resolver) adds 20--50\KB{} SRAM.  \tecompile{} eliminates this overhead
entirely by generating model-specific code with no interpreter loop.

\textbf{CMSIS-NN}~\cite{lai2018cmsis}.
ARM's hand-optimized kernel library for Cortex-M, commonly used as a backend
for TFLite Micro. It provides efficient \smlad{}-based convolution kernels.
\tecompile{} generates equivalent kernels from IR rather than linking a
pre-written library, enabling per-model specialization.

\textbf{ExecuTorch}~\cite{executorch2024}.
Meta's PyTorch-native edge deployment framework.  It supports MCUs via an
instruction-list interpreter (13--26\KB{} runtime) with pluggable
kernel backends. MCU support is demonstrated only on MNIST with CMSIS-NN delegation. 
In this architecture, \tecompile{} could serve as an ExecuTorch backend delegate for MCUs.

\textbf{Apache TVM / microTVM}~\cite{chen2018tvm}.
Compiler-based ML framework with emerging Cortex-M support. It still requires
a TVM C runtime with operator dispatch.  Auto-tuning (AutoTVM, MetaSchedule)
targets Cortex-A NEON or GPU architectures, not Cortex-M DSP instructions.

\textbf{IREE}~\cite{iree}.
Google's production MLIR compiler for ML workloads.  It targets GPUs, CPUs
(x86, ARM Cortex-A), and accelerators.  It requires an OS with threading and memory allocation; MCU targets are not supported.

\textbf{Glow}~\cite{rotem2019glow}.
Facebook's AOT neural network compiler.  It supports AOT compilation to C as \tecompile{} does and targets Cortex-A and Cortex-M, but via the support of CMSIS-NN.

\textbf{cardputer-ai}~\cite{cardputerai}.
A hand-written C++ LLM inference engine for the ESP32-S3 (Xtensa LX7,
8\MB{} PSRAM).  It demonstrates that LLM inference is feasible on MCUs, but the engine is hand-tuned for a single architecture.
\tecompile{} automates this: the same compiler generates LLM engines for multiple architectures from a standard model file.

\section{Conclusion}\label{sec:conclusion}

We presented \tecompile{}, the first complete MLIR-based compiler that
generates self-contained, dependency-free C inference engines for MCUs. The approach is demonstrated with the following results:

\textbf{Self-contained C code generation.}
For each input model, \tecompile{} emits all C artifacts needed for deployment: the inference
engine with all kernels and weights, the RTOS integration layer, a public API
header, and a CMake library target.  The output compiles with any cross-compiler and integrates into existing firmware in a few lines of code.  No
interpreter, no kernel library, no framework dependencies.  We believe this is the right
abstraction for MCU deployment: a complete, model-specific C library that a
firmware engineer can drop into the project.

\textbf{LLM inference on a single-core Cortex-M7.}
\tecompile{} compiles an 8.3M-parameter GPT-Neo (TinyStories-v2) to a
self-contained C engine that fits in 249\KB{} SRAM + 3.62\MB{} Flash and achieves $\sim$1\,token/s
interactive text generation on a \$25 NXP S32K344
automotive MCU.  The generated engine includes a stateful int4 KV cache with
a 96-token context window, enabling multi-turn conversation. To our knowledge, no existing MCU
inference framework supports this capability.

\textbf{Performance.}
On 10 MCUNet CNN benchmarks, \tecompile{} uses 38\% less SRAM and 14\%
less Flash than MCUNet TinyEngine, with identical ground-truth label
accuracy to TFLite on all models.

The key insight is that MLIR's multi-level dialect abstraction provides the
right infrastructure for MCU inference compilation: high-level enough to
express fusion patterns and memory planning, low-level enough to target DSP
instructions, and structured enough to emit complete, self-contained C code
rather than relying on pre-written kernel libraries.

\textbf{Limitations and future work.}
Testing and debugging \tecompile{} with a larger set of CNN and LLM models are in progress.
Lowering for Xtensa (ESP32-S3), AURIX Tricore, and their on-device performance measurements are left as future work. 

The residual 2--9 LSB error on deep networks with real inputs could potentially 
be reduced further by adopting TF/IREE’s approach of keeping \texttt{tosa.pad} as 
a separate op (avoiding the \texttt{Pad+Conv} folding complexity) or by verifying 
rounding at each intermediate layer.

\bibliographystyle{ACM-Reference-Format}
\bibliography{references}


\appendix

\section{Supplementary Material}\label{app:ops-passes}

This appendix provides usage examples for the \tecompile{} compiler
(Sections~\ref{app:pipeline-cmds}--\ref{app:suggest-output}),
followed by a complete reference of every MLIR operation
(Sections~\ref{app:te-ops}--\ref{app:target-ops}) and compilation pass
(Section~\ref{app:passes}) implemented by \tecompile{}.

\subsection{Compilation Pipeline Commands}\label{app:pipeline-cmds}

\tecompile{} accepts \texttt{.tflite}, \texttt{.safetensors}, and
\texttt{.mlir} files directly.  The examples below show the recommended
pass ordering for CNN and LLM models.

\paragraph{CNNs (int8 TFLite model, ARM Cortex-M target).}

\begin{scriptsize}
\begin{verbatim}
te-comp model.tflite \
  --convert-tosa-to-te \
  --te-flash-weights \
  --te-in-place-depthwise \
  --te-operator-specialize \
  --te-weight-prepack \
  --te-patch-inference \
  --te-memory-plan \
  --convert-te-to-loops \
  --convert-arm-dsp-to-llvm \
  --emit-c --rtos baremetal \
  -o output/
\end{verbatim}
\end{scriptsize}

\noindent
This produces a self-contained C inference engine in \textit{output/} directory:
\begin{itemize}[nosep]
\item The engine source (\textit{*\_engine.c/.h})
\item RTOS integration (\textit{inference\_task.c}, \textit{inference\_api.h})
\item A CMake target (\textit{CMakeLists.txt})
\item A machine-readable manifest (\textit{manifest.json})
\end{itemize}

\paragraph{LLMs (SafeTensors model, ESP32-S3 dual-core target).}

\begin{scriptsize}
\begin{verbatim}
te-comp model.safetensors \
  --sf-config config.json \
  --sf-seq-len 1 \
  --te-flash-weights \
  --te-kv-cache \
  --te-quantize-weights \
  --te-parallel-matmul \
  --convert-tosa-to-te \
  --te-memory-plan \
  --convert-te-to-loops \
    --target=xtensa \
  --convert-xtensa-to-llvm \
  --emit-c --rtos freertos \
  -o output/
\end{verbatim}
\end{scriptsize}

\noindent
The LLM-specific passes (\textit{--te-kv-cache}, \textit{--te-quantize-weights},
\textit{--te-parallel-matmul}) must precede \textit{--convert-tosa-to-te}.
In particular, \textit{--te-kv-cache} must run before
\textit{--te-quantize-weights} so that the position-embedding gather
pattern is recognized before weight quantization transforms it.

\paragraph{Auto-optimize (constraint-driven).}

\begin{scriptsize}
\begin{verbatim}
te-comp model.tflite \
  --auto-optimize \
  --sram-limit 128K \
  --flash-limit 512K \
  --emit-c -o output/
\end{verbatim}
\end{scriptsize}

\noindent
\texttt{--auto-optimize} runs the AOT analysis, then applies all
non-SKIP recommended passes automatically before proceeding to
code generation.

\paragraph{End-to-end pipeline summary.}
Table~\ref{tab:pipeline-summary} lists the 14 compilation steps in
recommended order, indicating which steps are CNN- or LLM-specific.

\subsection{Automated Optimization Suggestion}\label{app:suggest-output}

The \texttt{--te-suggest-optimizations} flag runs an AOT analysis that
inspects the TOSA IR and produces a prioritized recommendation report.
Example invocation and abbreviated output for the MCUNet-10fps-ImageNet
model on a NUCLEO-G474RE board (128\,KB SRAM, 512\,KB Flash):

\begin{scriptsize}
\begin{verbatim}
te-comp model.tflite \
  --te-suggest-optimizations \
  --sram-limit 128K --flash-limit 512K
\end{verbatim}
\end{scriptsize}

\begin{scriptsize}
\begin{verbatim}
Model type:    CNN (int8)
Total ops:     506 TOSA ops
Conv2d:        31 (30 pointwise 1x1, 1 spatial)
Depthwise:     15

Est. peak SRAM:    187.5 KB (budget: 128.0 KB -> OVER)
Est. weight flash: 787.1 KB (budget: 512.0 KB -> OVER)

Recommended optimizations:

[HIGH] --te-in-place-depthwise
       15 depthwise convs; peak SRAM over budget
       Saving: 1.4 MB | Accuracy loss: 0pp

[HIGH] --te-memory-plan
       static arena allocation reduces peak SRAM
       Accuracy loss: 0pp

[MED]  --te-operator-specialize
       30 pointwise 1x1 convs, 31 total conv2d
       Accuracy loss: 0pp

[MED]  --te-patch-inference
       largest activation exceeds half SRAM budget
       Saving: 93.8 KB | Accuracy loss: ~0.2pp

[HIGH] --te-flash-weights
       weights: 805988 B; >70% of flash budget
       Accuracy loss: 0pp

[SKIP] --te-kv-cache (no attention patterns)
[SKIP] --te-quantize-weights (already int8)
[SKIP] --te-parallel-matmul (single-core)

Aggregate accuracy loss: ~0.2pp
\end{verbatim}
\end{scriptsize}

\noindent
The analysis runs three phases: (1)~model signature extraction (op
inventory, model type detection, attention patterns, weight types),
(2)~resource estimation (peak activation SRAM, weight Flash, scratch
buffers), and (3)~cross-referencing the signature and estimates against
the target constraints to produce prioritized recommendations
(\texttt{HIGH}/\texttt{MED}/\texttt{LOW}/\texttt{SKIP}) with estimated
savings and accuracy impact.

\begin{table}[t]
  \centering
  \caption{End-to-end compilation pipeline in recommended order.
    Steps marked \textbf{L} are LLM-only; \textbf{C} are CNN-only;
    unmarked steps apply to both.}
  \label{tab:pipeline-summary}
  \scriptsize
  \setlength{\tabcolsep}{2pt}
  \begin{tabular}{@{}r p{3.7cm} p{3.4cm}@{}}
    \toprule
    \textbf{\#} & \textbf{Step / Flag} & \textbf{Summary} \\
    \midrule
    1  & Input conversion
       & \texttt{.tflite}/\texttt{.safetensors} $\to$ TOSA MLIR (Python converters) \\
    2  & \texttt{--te-flash-weights}
       & Mark large constants as flash-resident (zero SRAM) \\
    3  & \texttt{--te-kv-cache}~\textbf{L}
       & Insert stateful KV cache; must precede weight quantization \\
    4  & \texttt{--te-quantize-weights}~\textbf{L}
       & Q4\_0 or int8 weight quantization \\
    5  & \texttt{--te-parallel-matmul}~\textbf{L}
       & Multi-core row splitting \\
    6  & \texttt{--convert-tosa-to-te}
       & TOSA $\to$ \tedialect{} with op fusion \\
    7  & \texttt{--te-in-place-depthwise} \textbf{C}
       & Alias DW conv output to input \\
    8  & \texttt{--te-operator-specialize} \textbf{C}
       & Kernel variant hints \\
    9  & \texttt{--te-weight-prepack}~\textbf{C}
       & Im2col layout + tiling \\
    10 & \texttt{--te-patch-inference}~\textbf{C}
       & Patch-based spatial splitting \\
    11 & \texttt{--te-memory-plan}
       & Static arena allocation \\
    12 & \texttt{--convert-te-to-loops}
       & \tedialect{} $\to$ \texttt{scf.for} + intrinsics \\
    13 & \texttt{--convert-*-to-llvm}
       & Target intrinsics $\to$ LLVM/arith \\
    14 & \texttt{--emit-c}
       & C code emission (+ RTOS glue) \\
    \bottomrule
  \end{tabular}
\end{table}

\subsection{\tedialect{} Dialect Operations}\label{app:te-ops}

Table~\ref{tab:all-te-ops} provides a brief description of each of the
34 operations in the \tedialect{} dialect.

\begin{table*}[t]
  \centering
  \caption{Complete reference of the 34 \tedialect{} dialect operations.
    All operations are prefixed with \texttt{te.} in MLIR assembly
    (e.g., \texttt{te.conv2d}).}
  \label{tab:all-te-ops}
  \scriptsize
  \setlength{\tabcolsep}{4pt}
  \begin{tabular}{@{}l r l p{7.8cm}@{}}
    \toprule
    \textbf{Category} & \textbf{\#} & \textbf{Operation} & \textbf{Description} \\
    \midrule
    \multirow{22}{*}{\rotatebox[origin=c]{90}{\textbf{CNN (22 ops)}}}
    & 1  & \texttt{conv2d}             & Quantized 2D conv with fused bias, per-channel requantize (mult+shift), and activation clamp. \\
    & 2  & \texttt{depthwise\_conv2d}  & Depthwise 2D conv (channel-wise filter). Supports \texttt{in\_place} flag for output-aliased execution. \\
    & 3  & \texttt{transpose\_conv2d}  & Quantized transposed (deconvolution) 2D conv with fused requantize and clamp. \\
    & 4  & \texttt{fully\_connected}   & Quantized dense layer: $\text{clamp}(\text{requant}(\mathbf{x}\mathbf{W}^T + \mathbf{b}))$. \\
    & 5  & \texttt{avg\_pool2d}        & Quantized average pooling with optional fused rescale. \\
    & 6  & \texttt{max\_pool2d}        & Max pooling (operates directly on int8 values; no rescale). \\
    & 7  & \texttt{add}                & Quantized element-wise add with per-input rescale (handles differing scales). \\
    & 8  & \texttt{mul}                & Quantized element-wise multiply with output rescale. \\
    & 9  & \texttt{max}                & Element-wise maximum of two tensors. \\
    & 10 & \texttt{min}                & Element-wise minimum of two tensors. \\
    & 11 & \texttt{clamp}              & Clamp to $[\min, \max]$ (implements ReLU, ReLU6, etc.). \\
    & 12 & \texttt{rescale}            & Per-channel or per-tensor requantization (mult+shift+zero-point). \\
    & 13 & \texttt{softmax}            & Quantized softmax via LUT with configurable $\beta$ mult/shift. \\
    & 14 & \texttt{sigmoid}            & Quantized sigmoid (logistic) via LUT. \\
    & 15 & \texttt{tanh}               & Quantized tanh via LUT. \\
    & 16 & \texttt{hard\_swish}        & Quantized hard-swish: $x \cdot \text{relu6}(x+3)/6$. \\
    & 17 & \texttt{leaky\_relu}        & Quantized leaky ReLU with configurable $\alpha$ (mult/shift). \\
    & 18 & \texttt{reshape}            & Zero-copy reshape (reinterprets buffer layout). \\
    & 19 & \texttt{transpose}          & Dimension permutation. \\
    & 20 & \texttt{concat}             & Concatenation of variadic inputs along a given axis. \\
    & 21 & \texttt{pad}                & Constant-value padding (spatial H/W). \\
    & 22 & \texttt{upsample}           & Nearest-neighbor upsampling by integer scale factors. \\
    \midrule
    \multirow{2}{*}{\textbf{RNN (2)}}
    & 23 & \texttt{svdf}               & SVD Filter: decomposes dense layer into feature and time components (KWS). \\
    & 24 & \texttt{lstm}               & Standard LSTM cell with forget, input, cell, and output gates. \\
    \midrule
    \multirow{9}{*}{\rotatebox[origin=c]{90}{\textbf{LLM (9)}}}
    & 25 & \texttt{matmul\_q4}         & Q4\_0 quantized matmul (4-bit weights, bf16 per-block scales). \\
    & 26 & \texttt{gather}             & Token embedding lookup; supports Q4\_0 flash-resident tables with on-the-fly dequant. \\
    & 27 & \texttt{layer\_norm}        & Fused single-pass LayerNorm ($\gamma(x{-}\mu)/\sigma{+}\beta$). \\
    & 28 & \texttt{gelu}               & Fused GELU activation (tanh approx.), replaces 6-op TOSA decomposition. \\
    & 29 & \texttt{rope}               & Rotary Position Embedding (LLaMA-family). \\
    & 30 & \texttt{kv\_cache\_write}   & Quantize and store K/V into persistent KV cache (int4/int8/fp32). \\
    & 31 & \texttt{kv\_cache\_attention} & Causal multi-head attention reading quantized KV cache $[0,\text{pos}]$. \\
    & 32 & \texttt{kv\_slide}          & Sliding-window KV cache eviction for unbounded generation. \\
    & 33 & \texttt{pos\_embed}         & Runtime position embedding lookup via KV cache position counter. \\
    \midrule
    \textbf{Special (1)}
    & 34 & \texttt{se\_block}          & Fused squeeze-and-excite (avg-pool$\to$FC$_1$$\to$ReLU$\to$FC$_2$$\to$sigmoid$\to$mul). \\
    \bottomrule
  \end{tabular}
\end{table*}

\subsection{Target Intrinsic Dialect Operations}\label{app:target-ops}

Each target architecture has a dedicated MLIR dialect with thin wrappers
over hardware-specific instructions.  The \texttt{convert-te-to-loops} pass
emits the matching dialect's operations based on the \texttt{--target} flag.
A subsequent \texttt{convert-<target>-to-llvm} pass lowers these to LLVM IR
or pure \texttt{arith} ops.

\begin{table}[t]
  \centering
  \caption{Per-target intrinsic dialect operations.}
  \label{tab:target-ops}
  \scriptsize
  \begin{tabular}{@{}l l p{3.8cm}@{}}
    \toprule
    \textbf{Dialect} & \textbf{Operation} & \textbf{Description / HW Instruction} \\
    \midrule
    \multirow{5}{*}{\texttt{arm\_dsp}}
    & \texttt{smlad}      & Dual 16$\times$16 signed MAC (SMLAD) \\
    & \texttt{sxtb16}     & Sign-extend bytes 0,2 to halfwords (SXTB16) \\
    & \texttt{sadd16}     & SIMD add of two halfword pairs (SADD16) \\
    & \texttt{pkhbt}      & Pack bottom/top halfwords (PKHBT) \\
    & \texttt{requantize} & Per-channel requant (SMULL + shift) \\
    \midrule
    \multirow{10}{*}{\texttt{xtensa}}
    & \texttt{ee\_vmulas\_s8}   & 16$\times$ int8 vector MAC (PIE) \\
    & \texttt{mac\_i32}         & Scalar i32 MAC (remainder lanes) \\
    & \texttt{requantize}       & Per-channel requant \\
    & \texttt{ee\_vldbc\_8}     & Broadcast-load byte to 16 lanes \\
    & \texttt{ee\_vld\_128\_ip} & 128-bit vector load with post-incr. \\
    & \texttt{ee\_vsr\_32}      & Shift right 32-bit lanes by SAR \\
    & \texttt{ee\_andq}         & Bitwise AND two Q registers \\
    & \texttt{ee\_vsubs\_s8}    & Vector saturating subtract int8 \\
    & \texttt{ee\_zero\_accx}   & Zero the ACCX accumulator \\
    & \texttt{rur\_accx\_0}     & Read ACCX low 32 bits \\
    \cmidrule{2-3}
    & \texttt{bf16\_to\_fp32}   & bf16 $\to$ fp32 (shift left 16) \\
    \midrule
    \multirow{3}{*}{\texttt{tricore}}
    & \texttt{madd}        & 32$\times$32 MAC with 64-bit intermediate \\
    & \texttt{pack}        & Pack two 16-bit values into 32-bit \\
    & \texttt{requantize}  & Per-channel requant (MUL.U/MADD) \\
    \midrule
    \multirow{2}{*}{\texttt{generic}}
    & \texttt{mac\_i32}    & Scalar i32 MAC (arith.muli + arith.addi) \\
    & \texttt{requantize}  & Per-channel requant (pure i64 arith) \\
    \bottomrule
  \end{tabular}
\end{table}

\subsection{Compilation and Optimization Passes}\label{app:passes}

Table~\ref{tab:all-passes} lists all 16 registered MLIR passes plus the
AOT analysis system. The passes are listed in recommended pipeline
order: LLM-specific pre-processing, TOSA$\to$te conversion,
MCUNet optimizations, memory planning, kernel lowering, target lowering.

\begin{table*}[t]
  \centering
  \caption{Complete reference of all compilation and optimization passes
    in \tecompile{}, listed in recommended pipeline order.}
  \label{tab:all-passes}
  \small
  \begin{tabular}{@{}r l l p{7.5cm}@{}}
    \toprule
    \textbf{\#} & \textbf{Pass flag} & \textbf{Category} & \textbf{Description} \\
    \midrule
    \multicolumn{4}{@{}l}{\textit{LLM pre-processing passes (run before TOSA$\to$te conversion)}} \\
    1  & \texttt{--te-flash-weights}
       & LLM / CNN
       & Marks constants above a size threshold with \texttt{te.flash\_resident};
         memory planner skips them (zero SRAM), C emitter places them in \texttt{.rodata}. \\
    2  & \texttt{--te-kv-cache}
       & LLM
       & Detects attention patterns (reshape $[B,S,D] \to [B,S,n_h,d_h]$) and inserts
         \texttt{te.kv\_cache\_write} / \texttt{te.kv\_cache\_attention} ops for stateful
         autoregressive generation.  Replaces static wpe with \texttt{te.pos\_embed}. \\
    3  & \texttt{--te-quantize-weights}
       & LLM
       & Quantizes fp32 weight matrices to Q4\_0 (4-bit, bf16 scales, $\sim$7$\times$
         compression) or int8 ($\sim$4$\times$).  Reports per-layer RMSE/SNR at compile time. \\
    4  & \texttt{--te-parallel-matmul}
       & LLM
       & Annotates large matmuls with row-split attributes (\texttt{te.parallel\_split},
         \texttt{te.core\_ranges}) for multi-core execution.
         Supports equal and asymmetric (core-weights) splitting. \\
    \midrule
    \multicolumn{4}{@{}l}{\textit{Dialect conversion}} \\
    5  & \texttt{--convert-tosa-to-te}
       & Conversion
       & Lowers TOSA ops to \tedialect{} ops with pattern-based fusion:
         conv$+$rescale$+$clamp $\to$ \texttt{te.conv2d};
         5-op normalization $\to$ \texttt{te.layer\_norm};
         6-op activation $\to$ \texttt{te.gelu}; residual add detection. \\
    \midrule
    \multicolumn{4}{@{}l}{\textit{MCUNet optimization passes}} \\
    6  & \texttt{--te-in-place-depthwise}
       & MCUNet
       & Sets \texttt{in\_place} on depthwise conv2d ops where stride$\,=\,$1 and
         same spatial dims; memory planner aliases output to input. \\
    7  & \texttt{--te-operator-specialize}
       & MCUNet
       & Annotates conv/FC ops with kernel variant hints:
         pointwise 1$\times$1, IC/OC unrolling factors, small depthwise kernel. \\
    8  & \texttt{--te-weight-prepack}
       & MCUNet
       & Marks conv2d for im2col-based lowering and computes tiling hints
         (\texttt{te.tile\_oh}, \texttt{te.tile\_ow}) to fit L1 cache. \\
    9  & \texttt{--te-patch-inference}
       & MCUNet
       & Splits large conv2d inputs into overlapping spatial patches to
         reduce peak activation SRAM (patch-based inference). \\
    10 & \texttt{--te-bias-offset-fusion}
       & MCUNet
       & Folds the input zero-point$\times$weight-sum offset into the bias tensor
         at compile time, eliminating a runtime subtraction per output element. \\
    \midrule
    \multicolumn{4}{@{}l}{\textit{Memory planning}} \\
    11 & \texttt{--te-memory-plan}
       & Transform
       & Static arena allocation: lifetime analysis $\to$ in-place aliasing $\to$
         FirstFit-decreasing bin packing.  Annotates each op with \texttt{arena\_offset}.
         Packs model I/O into the same arena. \\
    \midrule
    \multicolumn{4}{@{}l}{\textit{Kernel lowering}} \\
    12 & \texttt{--convert-te-to-loops}
       & Lowering
       & Lowers all \tedialect{} ops to \texttt{scf.for} loops on
         \texttt{memref} buffers, emitting target-specific intrinsic ops
         based on \texttt{--target=\{arm,xtensa,tricore,generic\}}. \\
    \midrule
    \multicolumn{4}{@{}l}{\textit{Target-specific lowering to LLVM / arith}} \\
    13 & \texttt{--convert-arm-dsp-to-llvm}
       & Lowering
       & Lowers \texttt{arm\_dsp.*} ops to LLVM intrinsics
         (\texttt{llvm.arm.smlad}, \texttt{llvm.arm.sxtb16}, etc.)
         or pure \texttt{arith} patterns. \\
    14 & \texttt{--convert-xtensa-to-llvm}
       & Lowering
       & Lowers \texttt{xtensa.*} ops to Xtensa PIE assembly or
         library calls. \\
    15 & \texttt{--convert-tricore-to-llvm}
       & Lowering
       & Lowers \texttt{tricore.*} ops to \texttt{arith} patterns
         that the TriCore backend selects as MADD/PACK instructions. \\
    16 & \texttt{--convert-generic-to-llvm}
       & Lowering
       & Lowers \texttt{generic.*} ops to pure \texttt{arith} ops
         (portable fallback). \\
    \midrule
    \multicolumn{4}{@{}l}{\textit{Analysis (non-transforming)}} \\
    -- & \texttt{--te-suggest-optimizations}
       & Analysis
       & AOT model analysis: extracts model signature (op types, weight formats,
         attention dimensions), estimates memory, and produces a prioritized
         recommendation report (HIGH/MED/LOW/SKIP) with estimated savings. \\
    -- & \texttt{--auto-optimize}
       & Analysis
       & Runs the suggestion analysis, then applies all non-SKIP recommended
         passes automatically. \\
    \bottomrule
  \end{tabular}
\end{table*}

\subsection{C Code Emission}\label{app:emit-c}

In addition to the MLIR passes above, \tecompile{} includes a C code
emission backend (\texttt{--emit-c}) that translates the fully-lowered MLIR
(after \texttt{te-to-loops} and target-to-LLVM passes) into standalone C
source files.  The emitter handles: \texttt{arith.*} $\to$ C arithmetic;
\texttt{scf.for/if} $\to$ C loops/conditionals;
\texttt{memref.load/store} $\to$ flat array indexing;
\texttt{memref.alloca/view} $\to$ local arrays and arena pointer offsets;
\texttt{memref.get\_global} $\to$ \texttt{.rodata} constant references.
An optional \texttt{--rtos=\{freertos,baremetal\}} flag invokes a Python
template generator that produces RTOS glue code (task wrapper, API header,
CMake target) from the manifest JSON.

\noindent
The complete pipeline ordering is summarized in
Table~\ref{tab:pipeline-summary}.

\end{document}